\documentclass{aa}

\input psfig.sty

\begin{document}

\title{Evolutionary Population Synthesis for Binary Stellar Populations}

\author{Fenghui Zhang\inst{1}
        \and Zhanwen Han\inst{1}
        \and Lifang Li\inst{1}
        \and Jarrod R. Hurley\inst{2}
        }

\offprints{F. Zhang}

\institute{National Astronomical Observatories/Yunnan Observatory,
Chinese Academy of Sciences,
           Kunming, 650011, P.R. China\\
           \email{gssephd@public.km.yn.cn}
          \and
          Department of Astrophysics, American Museum of Natural History,
          Central Park West at 79th Street, New York, NY 10024, USA
          }

\date{Received date/ accepted date}

\abstract{We present integrated colours, integrated spectral
energy distributions, and absorption-line indices, for
instantaneous burst solar-metallicity binary stellar populations
with ages in the range $1 - 15\,$Gyr. By comparing the results for
populations with and without binary interactions we show that the
inclusion of binary interactions makes the appearance of the
population substantially bluer -- this is the case for each of the
quantities we have considered. This effect raises the derived age
and metallicity of the population. Therefore it is necessary to
consider binary interactions in order to draw accurate conclusions
from evolutionary population synthesis work.

 }

\titlerunning{Evolutionary Binary Stellar Populations}

\authorrunning{F. Zhang et al.}

\maketitle

\section{Introduction}

The majority of current studies of the integrated light of stellar
populations using the evolutionary population synthesis (EPS)
method -- the most direct approach for modelling galaxies -- have
tended to focus solely on the evolution of single stars. However,
observations tell us that upwards of 50\% of the stars populating
galaxies are expected to be in binary or higher-order multiple
systems (Duquennoy \& Mayor 1991; Richichi et al. 1994, for
example). Binary evolution, if the component stars are close
enough to exchange mass, can drastically alter the evolution path
of a star as expected from single star evolution. Binary
interactions can also create some important classes of objects,
such as blue stragglers (BSs: Pols \& Marinus 1994), and subdwarf
B stars (sdBs, also referred as extreme horizontal branch [EHB]
stars: Han et al. 2002; 2003). Therefore binary stars have the
potential to play an important role in determining the overall
appearance of any realistic stellar population and their evolution
should be accounted for in population synthesis models. A few EPS
studies have previously made advances in this direction (e.g. Pols
\& Marinus 1994; Cervino et al. 1997) but only for specialised
cases and not to the degree that we present here.

In this letter we assume that all stars are born in binaries and
born at the same time, i.e. an instantaneous binary stellar
population (BSP). We then model any interactions within these
binaries in our EPS models to investigate the effects on the
integrated colours, integrated spectral energy distributions
(ISEDs), and absorption feature indices. The outline of the paper
is as follows: we describe our EPS models and algorithm in Sect.
2, present our results and some discussion in Sect. 3, and then
finally in Sect. 4 we give our conclusions.

\section{Model Description}

\subsection{Input physics}

We use the rapid binary star evolution (BSE) algorithm of Hurley
et al. (2002) for the binary evolutionary tracks, the empirical
and semi-empirical calibrated BaSeL-2.0 model of Lejeune et al.
(1997; 1998) for the library of stellar spectra and the empirical
fitting functions of Worthey et al. (1994) for the spectral
absorption feature indices defined by the Lick Observatory image
dissector scanner (referred to as Lick/IDS) system.

The BSE algorithm provides the stellar luminosity $L$, effective
temperature $T_{\rm eff}$, radius $R$, current mass $M$ and the
ratio of radius to Roche-lobe radius $R/R_{\rm L}$ for the
component stars, as well as the period $P$, separation $a$ and
eccentricity $e$ for a binary system. It is valid for component
star masses in the range $0.1 \leq M_1, M_2 \leq 100 M_\odot$,
metallicity $0.0001 \leq Z \leq 0.03$, and eccentricity $0.0 \leq
e < 1.0$. The algorithm includes the single star evolution (SSE)
package of analytic formulae as presented by Hurley et al. (2000)
in its entirety. In fact, for orbits that are wide enough that
mass exchange between the component stars does not take place, the
evolutionary parameters of the stars are identical to that given
by the SSE package. In addition to all aspects of single star
evolution, the BSE algorithm models processes such as mass
transfer, mass accretion, common-envelope evolution, collisions,
supernova kicks, tidal evolution, and all angular momentum loss
mechanisms. This is done mostly by using a prescription (or
recipe) based approach.

The BaSeL-2.0 stellar spectra library of Lejeune et al. (1997;
1998) provides an extensive and homogeneous grid of low-resolution
theoretical flux distributions in the range of $9.1 - 160000\,$nm,
and synthetic UBVRIJHKLM colours for a large range of stellar
parameters: $2000 \leq T_{{\rm eff}}/ {\rm K} \leq 50000$, $-1.02
\leq \log g \leq 5.50$, and $+1.0 \leq [Fe/H] \leq -5.0$ (where
$g$ denotes surface gravity). For this library correction
functions have been calculated for each value of the $T_{{\rm
eff}}$ and for each wavelength in order to yield synthetic
UBVRIJHKLM colours matching the empirical colour-$T_{{\rm eff}}$
calibrations derived from observations at solar metallicity.
Semi-empirical calibrations for non-solar abundances ($[Fe/H] =
-3.5$ to +1.0) have also been established for this version of the
library. After correction the most important systematic
differences existing between the original model spectra and the
observations are eliminated. Furthermore, synthetic UBV and
Washington ultraviolet excesses $\delta_{(U-B)}$, $\delta_{(C-M)}$
and $\delta_{(C-T_1)}$, obtained from the original model spectra
of giants and dwarfs, are in excellent agreement with the
empirical metal-abundance calibrations.

The empirical fitting functions of Worthey et al. (1994) give
Lick/IDS absorption-line indices as a function of $T_{{\rm eff}}$,
$\log g$, and metallicity $[Fe/H]$. The effective temperature
spans a range of $2100 \leq T_{{\rm eff}}/{\rm K} \leq 11000$ and
the metallicity is in the range $-1.0 \leq [Fe/H] \leq +0.5$.
The indices in the Lick
system were extracted from the spectra of 460 stars obtained
between 1972 and 1984 using the red-sensitive IDS and Cassegrain
spectrograph on the 3m Shane telescope at Lick Observatory. The
spectra cover the range $4000-6400$ \AA, with a resolution of
$\sim 8$ \AA \ (Worthey et al. 1994).

\subsection{Model input and Monte Carlo simulations}

For the EPS of a BSP the main input model parameters are: (i) the
initial mass function (IMF) of the primaries; (ii) the lower and
upper mass cut-offs $M_{{\rm l}}$ and $M_{{\rm u}}$ to the IMF;
(iii) the mass-ratio distribution of the binaries; (iv) the
distribution of orbital separations; (v) the eccentricity
distribution; (vi) the relative age, $\tau$, of the BSP; and (vii)
the metallicity $Z$ of the stars.

Using the Monte Carlo method we simulate the real populations by
producing $2 \times 10^5$ binary systems. For each binary system
the initial mass of the primary is chosen from the approximation
to the IMF of Miller \& Scalo (1979) as given by Eggleton et al.
(1989), i.e.
\begin{equation}
M_1 = {\frac{ 0.19X }{(1-X)^{0.75}+0.032(1-X)^{0.25}}} \, ,
\label{mdis}
\end{equation}
where $X$ is a random variable uniformly distributed in the range
[0,1], and $M_1$ is the primary mass in unit of $M_\odot$. For
each BSP the lower and upper mass cut-offs $M_{{\rm l}}$ and
$M_{{\rm u}} $ are taken as 0.1 $M_{\odot}$ and 100 $M_{\odot}$.

The initial masses of the component stars are assumed to be
correlated with the initial mass of the secondary obtained from a
uniform mass-ratio distribution (Mazeh et al. 1992; Goldberg \&
Mazeh 1994), i.e.
\begin{equation}
q = X
\label{qdis}
\end{equation}
where $q = M_2/M_1$ is the mass-ratio.

The distribution of separations is taken as constant in $\log a$
for wide binaries and falls off smoothly at close separations:
\begin{equation}
a n(a) = \Bigl\{\matrix{\ a_{\rm sep}(a/a_0)^m , & \ \ \ a \leq
a_0, \cr a_{\rm sep}, \ \ \ \ \ \ \ \ \ & \ \ \ \ \ \ \ \ \ \  a_0
< a < a_1, \cr}
\label{adis}
\end{equation}
where $a_{\rm sep}\approx 0.070$, $a_0=10R_{\odot }$,
$a_1=5.75\times 10^6R_{\odot }$ and $m\approx 1.2$. This
distribution implies that there are equal numbers of wide binary
systems per logarithmic interval, and that approximately 50\% of
stellar systems are binary systems with orbital periods less than
100 yr (The fraction is a typical value for the Galaxy, resulting
in $\sim 10.1\%$ of the binaries experiencing Roche lobe overflow
during the past 13 Gyr, see Han et al. 1995).

We allow eccentric orbits for the binary systems and assume a
uniform eccentricity distribution, i.e.
\begin{equation}
e = X.  \label{edis}
\end{equation}

In our BSPs we assume that all stars are born in binaries in an
instantaneous burst of star formation and we vary the age of the
BSPs from $1 - 15\,$Gyr. The metallicity is taken as solar ($Z =
0.02$) throughout this work. For the BSE code there are several
important input parameters that require mention: the efficiency of
common envelope ejection $\alpha_{{\rm CE}}$ is taken as 1.0; the
Reimers wind mass-loss coefficient $\eta$ is set constant at 0.3;
and the tidal enhancement parameter $B = 0.0$.

\subsection{Algorithm}

Once the initial state of a binary system (the masses of the
component stars, $M_{{\rm 1}}$ and $M_{{\rm 2}}$, separation $a$
and eccentricity of the orbit $e$) is given, we obtain
evolutionary parameters such as luminosity, effective temperature,
radius, and mass for the component stars using the BSE algorithm,
transform these evolutionary parameters to colours and stellar
flux with the BaSeL-2.0 stellar spectral model, and obtain
absorption feature indices of the Lick/IDS system using the
fitting functions of Worthey et al. (1994). By the following
equations (\ref{inte-c} - \ref{inte-mag}) we can obtain the
integrated colours, monochromatic flux and absorption feature
indices for an instantaneous BSP of a particular age and
metallicity.

In the following equations, a parameter identified by a capital
letter on the left-hand side represents the integrated BSP, while
the corresponding parameter in minuscule on the right-hand side is
for the $k-$th star. The integrated colour is expressed by
\begin{eqnarray}
(C_{{\rm i}}-C_{{\rm j}})_{\tau, Z} & = -2.5 &
\log{\frac{\sum_{{\rm k=1}}^{ {\rm n}}\ 10^{-0.4c_{{\rm i}}}
}{\sum_{{\rm k=1}}^{{\rm n}}\ 10^{-0.4c_{{\rm j}}}}}
\label{inte-c}
\end{eqnarray}
where $c_{{\rm i}}$ and $c_{{\rm j}}$ are the $i$-th and $j$-th
magnitude of the $k$-th star.

The integrated monochromatic flux of a BSP is defined as
\begin{equation}
F_{\lambda,\tau,Z} = \sum_{{\rm k=1}}^{{\rm n}} \ f_{\lambda},
\label{sp-lamda}
\end{equation}
where $f_{\lambda}$ is the SED of the $k$-th star.

The integrated absorption feature index of the Lick/IDS system is
a flux-weighted one. For the $i-$th atomic absorption line, it is
expressed in equivalent width ($W$, in \AA),
\begin{equation}
W_{i,\tau,Z} = {\frac{\sum_{{\rm k=1}}^{{\rm n}} \ w_{{\rm i}} \
\cdot f_{i,{\rm C}\lambda} }{\sum_{{\rm k=1}}^{{\rm n}} \ f_{i,{\rm
C}\lambda}}},
\label{inte-EW}
\end{equation}
where $w_{{\rm i}}$ is the equivalent width of the $i-$th index of the $k$-th
star, and $ f_{i,{\rm C} \lambda}$ is the continuum flux at the
midpoint of the $i-$th `feature' passband; and for the $i-$th
molecular line, the feature index is expressed in magnitude,
\begin{equation}
C_{i,\tau,Z} = -2.5 \ {\rm log} {\frac{\sum_{{\rm k=1}}^{{\rm
n}} \ 10^{-0.4 c_{{\rm i}}} \cdot \ f_{i,{\rm C}\lambda} }{\sum_{{\rm
k=1}}^{{\rm n}} \ f_{i,{\rm C}\lambda}}},  \label{inte-mag}
\end{equation}
where $c_{{\rm i}}$ is the magnitude of the $i-$th index of the
$k$-th star (as in equation~\ref{inte-c}).

\section{Results and Discussion}

\begin{figure}
\psfig{file=H4722F1.ps,height=6.0cm,bbllx=526pt,bblly=127pt,bburx=136pt,bbury=626pt,clip=,angle=270}
\caption{Theoretical isochrones for solar-metallicity
instantaneous burst BSPs at an age of 1 Gyr (solid circles for the
primary, open for the secondary). The left panel represents
Model~A (binary interactions are considered) and the right panel
represents Model~B (neglecting binary interactions). Blue
stragglers and extreme horizontal branch stars are clearly evident
in the left panel. For the sake of clarity only $5 \times 2^4$
binary systems are included in each panel.} \label{syn-b3}
\end{figure}

\begin{figure*}
\psfig{file=H4722F2.ps,height=12.0cm,bbllx=580pt,bblly=113pt,bburx=79pt,bbury=687pt,clip=,angle=270}
\caption{Integrated colours as a function of age for
solar-metallicity instantaneous burst BSPs with (solid) and
without (open) binary interactions.} \label{color}
\end{figure*}

\begin{figure}
\psfig{file=H4722F3.ps,height=6.0cm,bbllx=580pt,bblly=39pt,bburx=79pt,bbury=702pt,clip=,angle=270}
\caption{The integrated spectral energy distributions for
solar-metallicity instantaneous burst BSPs with (full line) and
without (dashed line) binary interactions at ages of 1 Gyr and 13
Gyr (upper lines and lower lines, respectively). The flux is
expressed in magnitude and is normalized to zero at $2.2 \,
\mu$m.} \label{ised}
\end{figure}

\begin{figure*}
\psfig{file=H4722F4.ps,height=12.0cm,bbllx=580pt,bblly=39pt,bburx=79pt,bbury=700pt,clip=,angle=270}
\caption{Evolution of absorption indices in the Lick/IDS system
for solar-metallicity instantaneous burst BSPs with (solid) and
without (open) binary interactions.} \label{asi}
\end{figure*}

We have constructed two distinct BSPs in order to investigate the
effects of binary interactions on the integrated colours, ISEDs
and Lick/IDS absorption feature indices for instantaneous burst
solar-metallicity BSPs with ages in the range $1\leq (\tau /{\rm
Gyr})\leq 15$: Model A includes binary interactions by utilising
the BSE algorithm whereas Model B neglects all binary
interactions, i.e. the component stars are evolved as if in
isolation according to the SSE algorithm.

In Fig. \ref{syn-b3} we show the theoretical isochrones for Models
A and B at an age of $\tau =1\,$Gyr. As expected, we see clearly
that the distribution of stars in the Hertzsprung-Russell diagram
for Model A is significantly different from that of Model B: (i)
the distribution of the stars is more dispersed for Model A in
comparison to Model B; and (ii) BSs and EHB stars on the helium
main-sequence are produced by Model A only. These differences in
the distribution of stars are responsible for differences in the
appearance of EPS models with and without binaries. As for old
BSPs (such as $\tau$= 13Gyr) the number of BSs in Model A is
relatively smaller, and no stars exist on the helium
main-sequence. Hurley et al. (2001) compared their theoretical
colour-magnitude diagram (CMD) with that of M67, it shows that our
treatment of binaries is realistic.

In Fig. \ref{color} we give a comparison of the integrated $U-B$,
$B-V$, $V-R$ and $V-I$ colours for Models A and B. We see that the
colours for Model A are bluer than those for Model B in all
instances, and at some ages the discrepancies become so large, for
example the difference in $B-V$ reaches $\sim 0.07\,$ at $\tau
=3\,$Gyr, that we cannot neglect the effects of binary
interactions in EPS models.

In Fig. \ref{ised} we give the ISEDs over a wide wavelength range,
$2.7\leq {\rm log}(\lambda /{\rm \AA })\leq 5.2$, for Model A and
Model B at ages of $1\,$Gyr and $13\,$Gyr. The flux is expressed
in magnitude and is normalized to zero at 2.2 $\mu $m. Fig.
\ref{ised} shows a significant disagreement of the ISED in the
far-ultraviolet region ($3.0 < {\rm log}(\lambda /{\rm \AA
})<3.2$) for $\tau = 1\,$Gyr BSPs, in that the ISED for Model A is
as much as $\sim 5\,$mag greater than that for Model B. The main
reason for this discrepancy is the existence of the hot BSs in
Model~A which dominate the ISED in this region. Also Fig.
\ref{ised} shows that the ISED for Model A exhibits bluer
continuum than that for Model B in the visible and infra-red
regions for $\tau = 1\,$Gyr BSPs -- this discrepancy is caused not
only by the existence of the hot BSs in Model~A but also by the
presence of giants whose colours have been altered by binary
interaction. For old BSPs (for example $\tau = 13\,$Gyr) the ISED
of Model A is distinguished from that of Model B only in the
wavelength range ${\rm log}(\lambda /{\rm \AA })<3.4$: it is bluer
than that for Model B by no more than $\sim 2\,$mag.

In Fig. \ref{asi} we show the evolution of two age-sensitive
Lick/IDS spectral absorption feature indices, G4300 and H$_{{\rm
\beta}}$, and two metallicity-sensitive indices, Fe5015 and
Mg$_{{\rm 2}}$, for Models A and B. Each of these indices are
bluer for Model A in comparison to Model B: the H$_{{\rm \beta}}$
indice for Model A is greater by $\sim$ 0.34 \AA \ at 3 Gyr.

From all the above comparisons, we can draw the conclusion that
binary interactions make the integrated colours, ISEDs and
Lick/IDS absorption feature indices bluer. As each of these
quantities tend to become redder with increasing age and
metallicity it follows that the inclusion of binary interactions
acts to raise the derived age and metallicity of a particular
population. The degree of this increase depends on the method, by
which one obtains the age and metallicity.

We also present the differences in the integrated colours and the
derived ages between Model A and B in Table \ref{colours}, and the
Lick/IDS indices in Table \ref{Licks}. At early and intermediate
ages ($\tau < 10\,$Gyr) the differences in the derived ages
($\Delta \rm{Age}$) reach the maximum because the number of BSs
peaks at $\sim 5\,$Gyr, and at late ages ($\tau < 10\,$Gyr) the
larger differences in the derived ages result from the facts (1)
the colours and Lick/IDS indices increase relatively lowly with
age, and (2) Monte Carlo simulations produce some fluctuations in
colours and indices, which will be improved in future studies.

\section{SUMMARY AND CONCLUSIONS}

We have simulated realistic stellar populations composed of 100\%
binaries by producing $2 \times 10^5$ binary systems using a Monte
Carlo technique. We computed the integrated colours, ISEDs and
Lick/IDS absorption feature indices for these instantaneous burst
BSPs with and without binary interactions. In comparison we find
that modelling binary evolution, and the additional classes of
stars that this produces, leads to bluer integrated colours, ISEDs
and Lick/IDS absorption feature indices, and therefore makes the
derived age and metallicity raise. In this letter we have only
considered the effects of binaries for solar metallicity BSPs --
more detailed studies will be given later.

\begin{acknowledgements}
We acknowledge the generous support provided by the Chinese
Natural Science Foundation (Grant No. 19925312, 10073009, 10273020
\& 10303006), Yunnan Natural Science Foundation (Grant No.
2002A0020Q) and by the 973 scheme (NKBRSF G1999075406). We are
deeply indebted to Dr. Lejeune for making his BaSeL-2.0 model
available to us. We thank an anonymous referee for his/her useful
suggestions.
\end{acknowledgements}

{}

\begin{table*}
 \caption []{The differences in the integrated colours and the derived ages
             between model A and B.}
   \begin{tabular}{ccccccccc}
    \hline\noalign{\smallskip}
     Age & $\Delta (U-B)$ & $\Delta \rm{Age}_{(U-B)}$
         & $\Delta (B-V)$ & $\Delta \rm{Age}_{(B-V)}$
         & $\Delta (V-R)$ & $\Delta \rm{Age}_{(V-R)}$
         & $\Delta (V-I)$ & $\Delta \rm{Age}_{(V-I)}$ \\
    (Gyr)& (mag) & (Gyr) & (mag) & (Gyr) &  (mag) & (Gyr) &  (mag) & (Gyr) \\
    \hline
    1.0 &  -0.032 &  0.865 &  -0.033 &  0.187 &  -0.012 &  0.118 &  -0.039 &  0.204 \\
    2.0 &  -0.033 &  1.500 &  -0.050 &  0.980 &  -0.022 &  0.710 &  -0.046 &  0.885 \\
    3.0 &  -0.084 &  0.706 &  -0.069 &  0.831 &  -0.025 &  0.893 &  -0.054 &  2.842 \\
    4.0 &  -0.033 &  0.465 &  -0.039 &  0.650 &  -0.021 &  0.656 &  -0.055 &  0.679 \\
    5.0 &  -0.022 &  1.100 &  -0.026 &  3.714 &  -0.018 &  3.600 &  -0.050 &  6.250 \\
    6.0 &  -0.033 &  0.892 &  -0.030 &  1.154 &  -0.013 &  1.083 &  -0.027 &  0.794 \\
    7.0 &  -0.041 &  1.519 &  -0.030 &  1.579 &  -0.016 &  0.842 &  -0.037 &  0.755 \\
    8.0 &  -0.041 &  1.449 &  -0.029 &  2.506 &  -0.013 &  3.500 &  -0.030 &  8.750 \\
    9.0 &  -0.046 &  1.626 &  -0.031 &  2.679 &  -0.015 &  4.038 &  -0.037 &  7.792 \\
   11.0 &  -0.066 &  2.333 &  -0.044 &  3.802 &  -0.026 &  7.000 &  -0.074 &  4.583 \\
   13.0 &  -0.052 &  1.838 &  -0.031 &  2.679 &  -0.012 &  3.231 &  -0.016 &  4.667 \\
   15.0 &  -0.052 &  2.849 &  -0.025 &  4.545 &  -0.008 &  4.000 &  -0.022 &  4.000 \\
   \hline
   \end{tabular}
   \label{colours}
\end{table*}

\begin{table*}
 \caption []{The difference in Lick/IDS indices and the derived ages
             between model A and B}
   \begin{tabular}{ccccccccc}
    \hline\noalign{\smallskip}
     Age & $\Delta \rm{G4300}$   & $\Delta \rm{Age}_{G4300}$
         & $\Delta \rm{H_\beta}$ & $\Delta \rm{Age}_{H_\beta}$
         & $\Delta \rm{Fe5015}$  & $\Delta \rm{Age}_{Fe5015}$
         & $\Delta \rm{Mg_2}$    & $\Delta \rm{Age}_{Mg_2}$ \\
    (Gyr)& ($\rm{\AA}$) & (Gyr) & ($\rm{\AA}$) & (Gyr)
         & ($\rm{\AA}$) & (Gyr) & (mag) & (Gyr) \\
    \hline
    1.0 &  -0.245 &  0.121 & 0.139 &  0.090 &  -0.187 &  0.222 &  -0.004 &  0.096 \\
    2.0 &  -0.449 &  1.121 & 0.252 &  0.778 &  -0.232 &  1.116 &  -0.010 &  0.498 \\
    3.0 &  -0.732 &  0.874 & 0.341 &  0.802 &  -0.313 &  1.049 &  -0.013 &  0.687 \\
    4.0 &  -0.324 &  0.797 & 0.168 &  0.852 &  -0.197 &  0.654 &  -0.011 &  0.722 \\
    5.0 &  -0.234 &  0.944 & 0.122 &  0.936 &  -0.154 &  8.067 &  -0.012 &  1.219 \\
    6.0 &  -0.205 &  0.866 & 0.107 &  0.964 &  -0.140 &  1.082 &  -0.008 &  0.925 \\
    7.0 &  -0.162 &  1.652 & 0.084 &  1.231 &  -0.165 &  1.681 &  -0.008 &  0.709 \\
    8.0 &  -0.172 &  0.818 & 0.072 &  0.690 &  -0.176 &  3.806 &  -0.007 &  1.219 \\
    9.0 &  -0.130 &  1.047 & 0.057 &  0.766 &  -0.185 &  3.296 &  -0.007 &  1.238 \\
   11.0 &  -0.134 &  6.781 & 0.042 &  5.954 &  -0.328 & 15.570 &  -0.010 &  4.703 \\
   13.0 &  -0.179 &  2.612 & 0.088 &  1.775 &  -0.119 &  4.646 &  -0.008 &  2.033 \\
   15.0 &  -0.149 &  2.164 & 0.055 &  1.102 &  -0.148 &  5.798 &  -0.005 &  1.315 \\
   \hline
   \end{tabular}
   \label{Licks}
\end{table*}

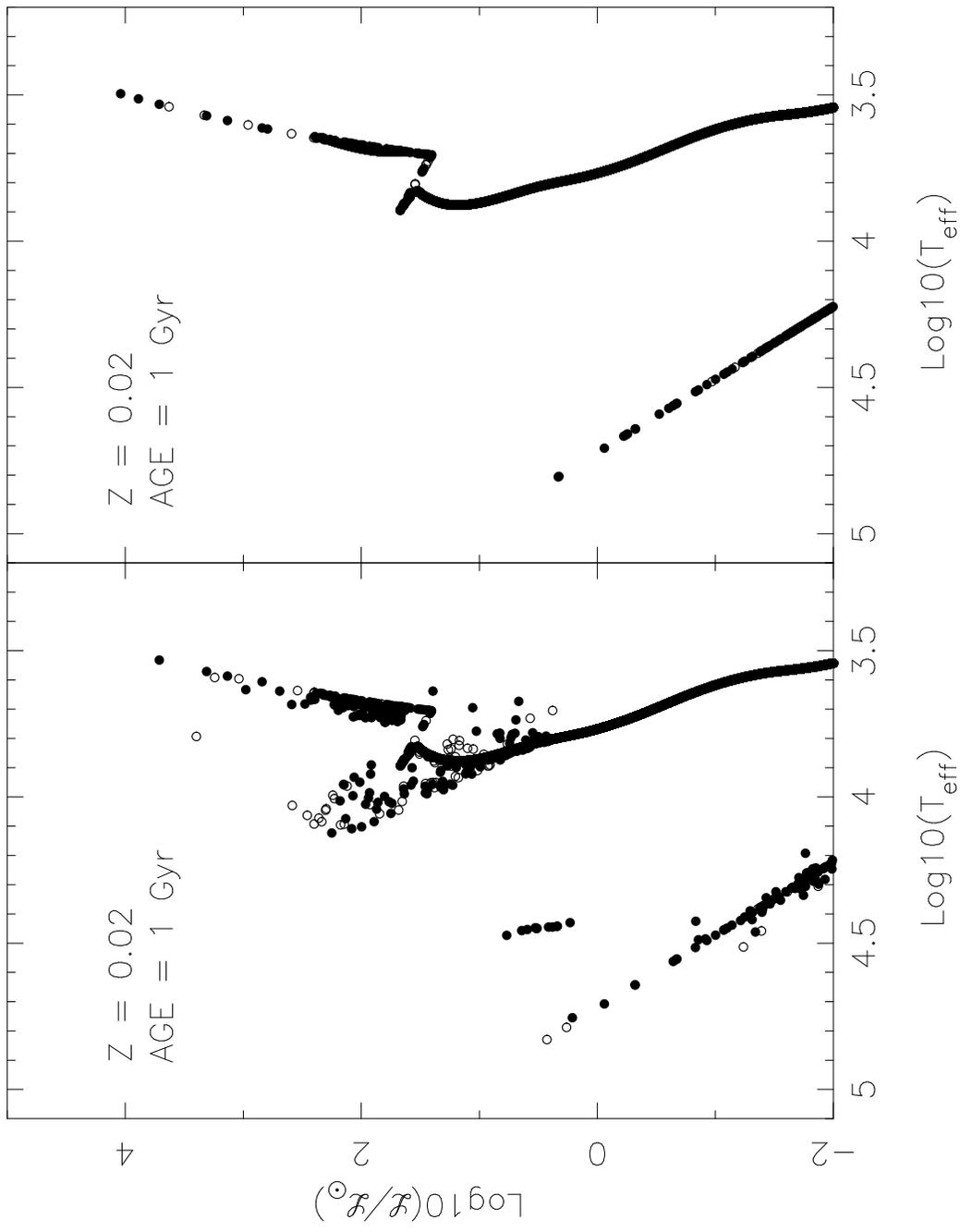
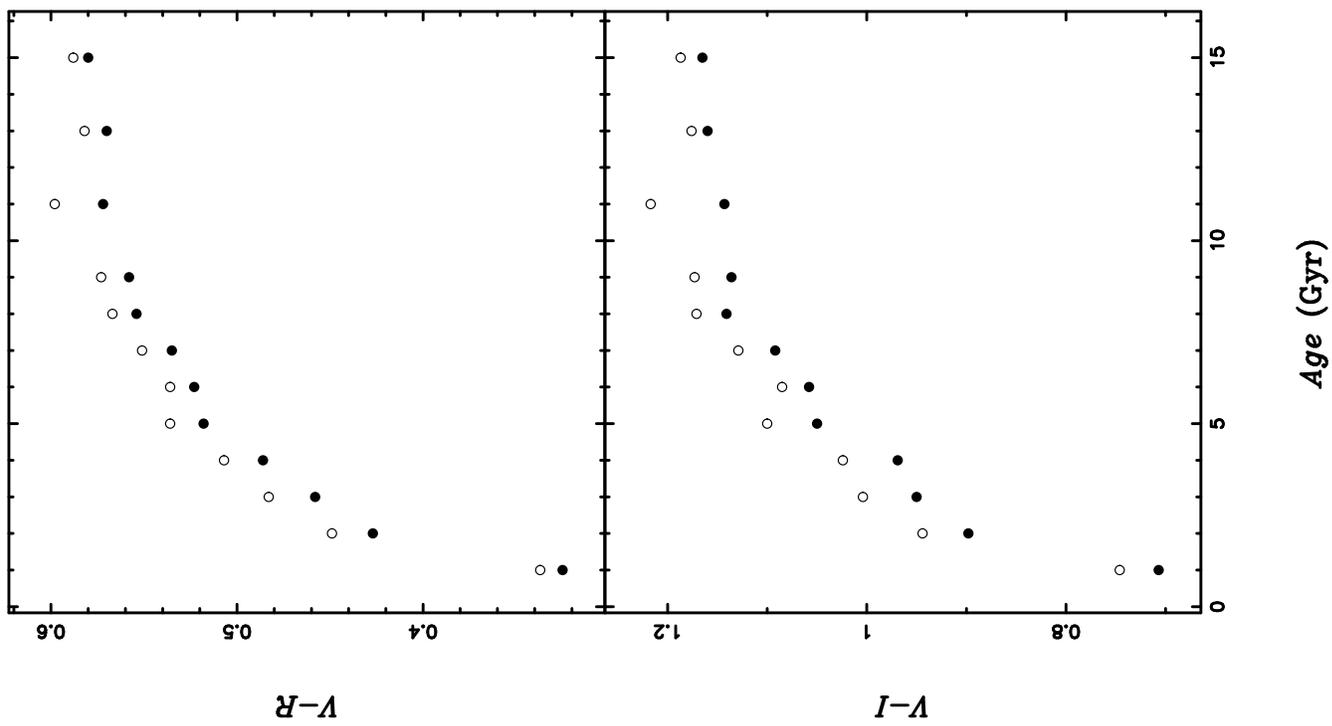
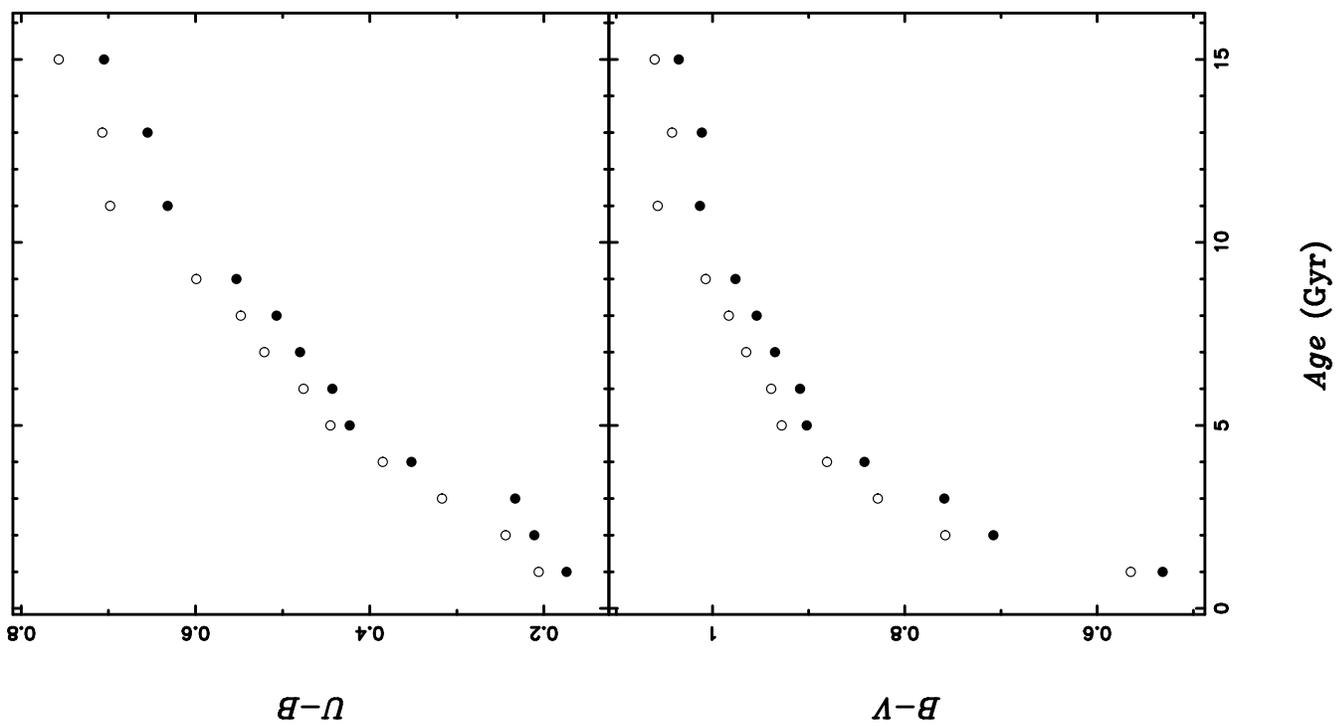
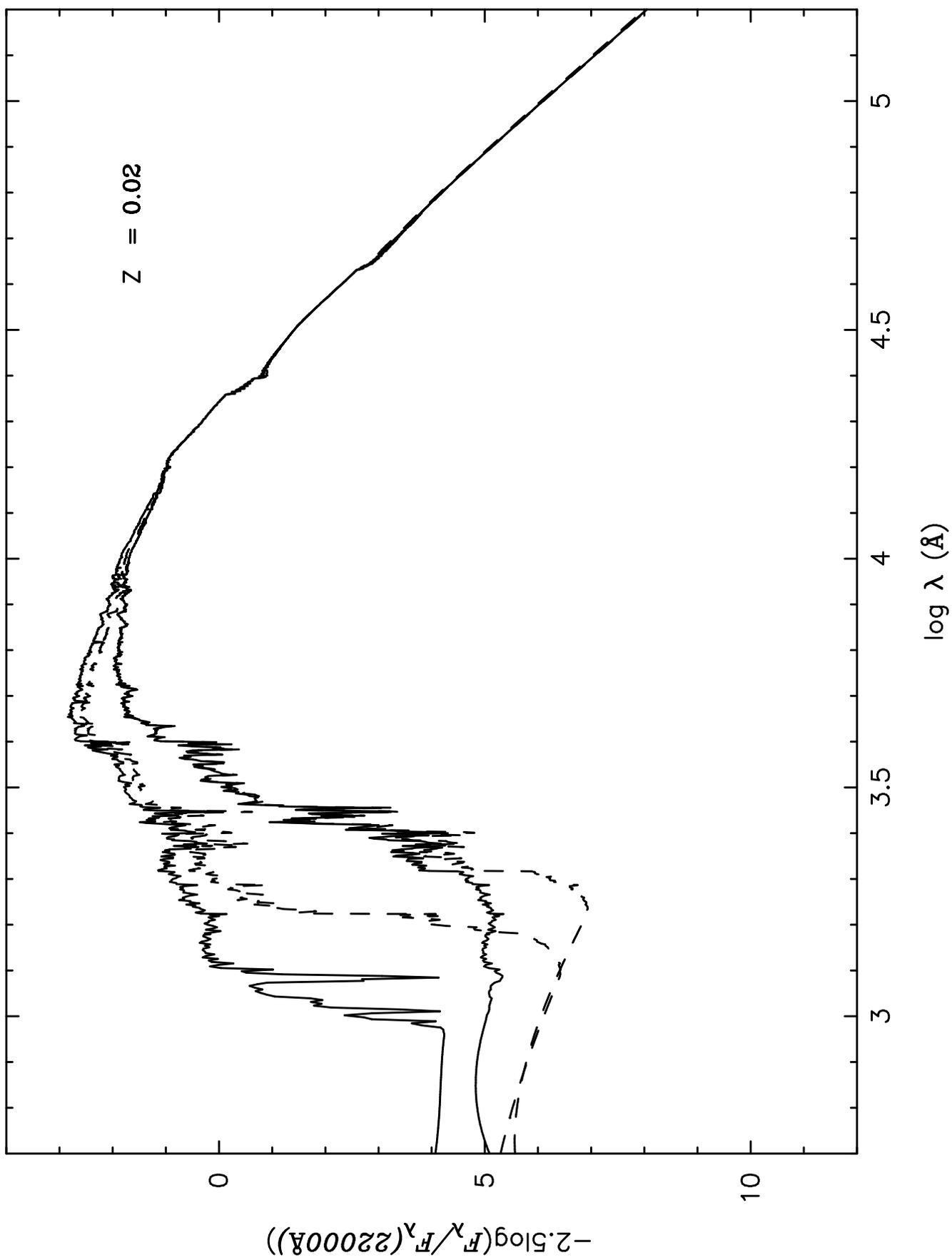
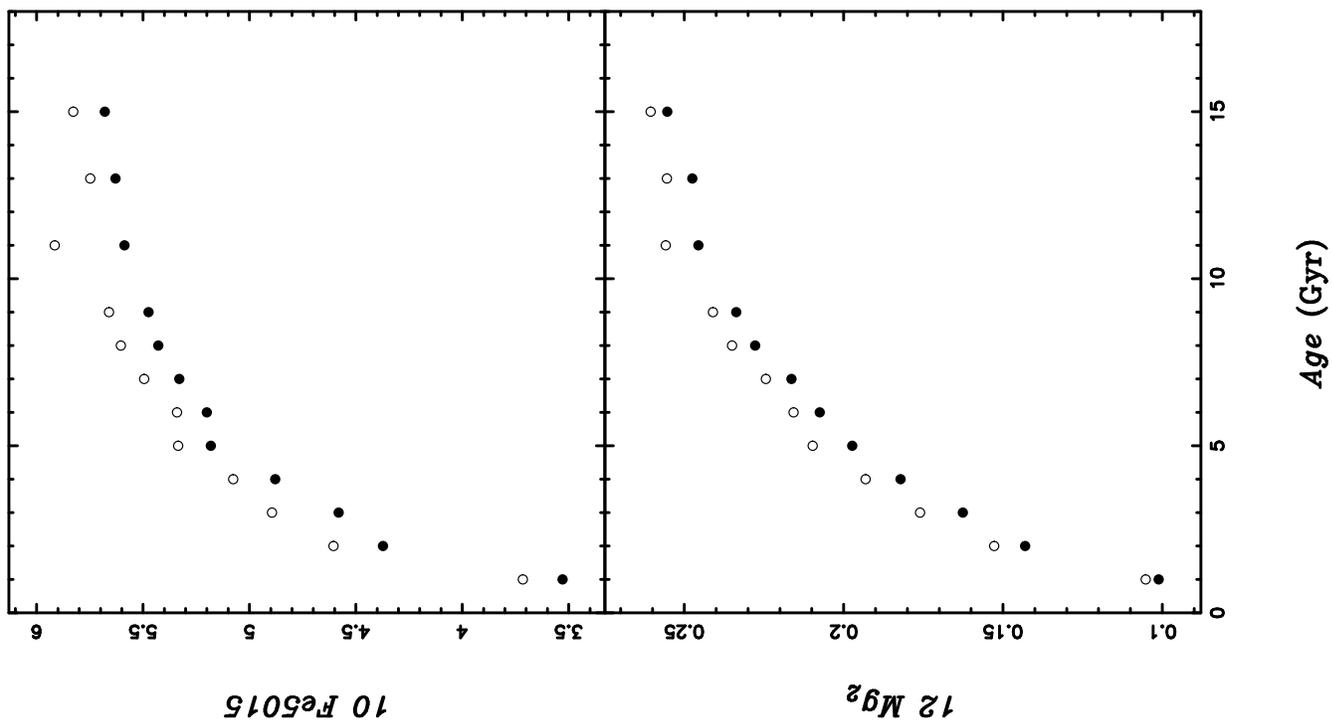
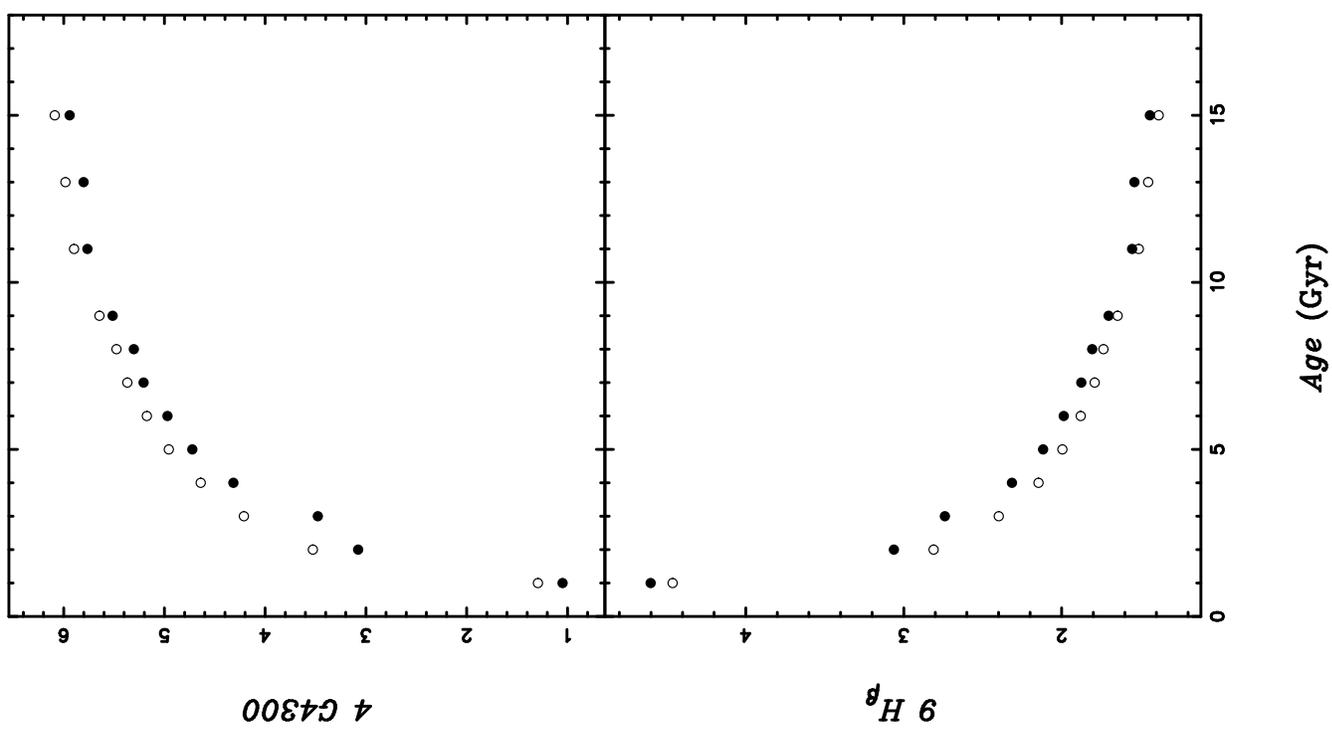
\end{document}